# Disentangling Rotational Dynamics and Ordering Transitions in a System of Self-Organizing Protein Nanorods *via* Rotationally Invariant Latent Representations

Sergei V. Kalinin,[1,1] Shuai Zhang,[2,3] Mani Valleti,[4]

Harley Pyles,[5,6] David Baker,[5,6,7] James J. De Yoreo,[2,3] and Maxim Ziatdinov[1,8,2]

[1] Center for Nanophase Materials Sciences, Oak Ridge National Laboratory, Oak Ridge, TN 37831, USA

[2] Materials Science and Engineering, University of Washington, Seattle, WA 98195, USA

[3] Physical Sciences Division, Pacific Northwest National Laboratory, Richland, WA 99354, USA

[4] Bredesen Center for Interdisciplinary Research, University of Tennessee, Knoxville, TN 37996, USA

[5] Department of Biochemistry, University of Washington, Seattle, WA 98195, USA

[6] Institute for Protein Design, University of Washington, Seattle, WA 98195, USA

[7] Howard Hughes Medical Institute, University of Washington, Seattle, WA 98195, USA

[8] Computational Sciences and Engineering Division, Oak Ridge National Laboratory, Oak Ridge, TN 37831, USA

**Abstract**

The dynamic of complex ordering systems with active rotational degrees of freedom exemplified by protein self-assembly is explored using a machine learning workflow that combines deep learning-based semantic segmentation and rotationally invariant variational autoencoder-based analysis of orientation and shape evolution. The latter allows for disentanglement of the particle orientation from other degrees of freedom and compensates for lateral shifts. The disentangled representations in the latent space encode the rich spectrum of local transitions that can now be visualized and explored *via* continuous variables. The time dependence of ensemble averages allows insight into the time dynamics of the system, and in particular, illustrates the presence of the potential ordering transition. Finally, analysis of the latent variables along the single-particle trajectory allows tracing these parameters on a single particle level. The proposed approach is expected to be universally applicable for the description of the imaging data in optical, scanning probe, and electron microscopy seeking to understand dynamics of complex systems where rotations are a significant part of the process.

---

[1] sergei2@ornl.gov
[2] ziatdinovma@ornl.gov

Emergence of ordered patterns in the systems of interacting particles is one of the foundational phenomena in chemistry,[1] condensed matter physics,[2] materials science,[3] and biology, encompassing areas ranging from the formation of atomic lattices, protein complexes, and lipid membranes, to the self-assembly of viruses and nanoparticles.[2, 4-6] Correspondingly, understanding of the evolution of such systems and the mechanisms guiding the emergence of the order have remained on the forefront of physical research for over half a century. This effort includes both the theoretical and simulation analysis,[7, 8] exploring both natural structures, and model systems such as colloidal crystals[2, 9] that allow for tunability of underpinning interactions.[10, 11]

One of the challenges in understanding the dynamics of pattern emergence in these systems is the nature of the descriptors of the systems, *i.e.* the compact representation of the local and global geometries. In the systems with long-range translation symmetry, like crystalline lattices, these can be naturally described in the form of the lattice and primitive unit cells. Deviations from ideal behaviors can be further analyzed in terms of the symmetry breaking phenomena, which yields the concepts of order parameters, *etc*. However, the descriptions are considerably more complex in the absence of long-range translational symmetry, as is the case for liquids and liquid crystals. The local descriptors are often constructed based on corresponding correlation functions and structure factors. The advantage of this approach is that it provides descriptors that can be derived from macroscopic scattering experiments, allowing experimental studies of such systems as a function of global stimuli such as temperature, chemical potentials, external fields, and time.

The rapid emergence of real space imaging methods and corresponding model systems have provided the insight into the mechanisms and dynamics of the self-organization phenomena on the single particle levels. Notable examples include optical microscopy of colloid crystals assembly, scanning probe microscopy of nanoparticle and molecular self-assembly,[12-14] and environmental electron microscopy of nanoparticle dynamics.[15, 16] As such, observations of system evolution particle-by particle and even the partial or full trajectory reconstructions have become common. The progress in the dynamic scanning transmission electron microscopy enabled these studies even on the atomic level, providing insight into the chemical reactions and phase transformations induced by the electron beam, temperature, and other environmental stimuli.[17-22]

However, this proliferation of the modelling and experimental data has brought forth the challenge of associated local descriptors, as a necessary step towards comparing local mechanisms to the theoretical models and scattering studies. For spherical particles interacting through isotropic or angle-dependent force fields, such local descriptors can be derived from local bonding geometries. Traditionally, these descriptors are based on the coefficients of the spherical functions expansions, giving rise to *e.g.* tetratic and hexatic order parameters.[23-25] Recent emergence of the machine learning methods, specifically linear dimensionality reduction methods such as principal component analysis, density based methods such as DBSCAN and UMAP, and manifold learning methods such as Isomap and tSNE have stimulated exploration of these systems using these reduced implementations.[26] Similarly, self-organized feature maps and (variational) autoencoders have recently been explored.[27] Common for all these methods is the projection of the high-dimensional local descriptors (*e.g.* coordinates of particle neighbors) on the low dimensional manifold, providing the reduced descriptors of local geometry that can further be correlated with functional properties. Recent overview of this field is given by Volpe.[28] In the last year, the advances in graph networks have stimulated rapid emergence of graph based methods for structural descriptoion.[29-31] Finally, when applied to time-dependent structural data, the machine learning methods have been used to identify collective variables[32-34] and reaction pathways.[35]

However, extension of this approach for the anisotropic particles, *e.g.* rod-like or having more complex geometries, leads to the rapid growth of the number of possible ad-hoc descriptors.[36, 37] The primary difficulty in this case becomes the presence of multiple rotational variants, leading to entanglement between particle shapes and inter particle geometric parameters such as spacings and interaction geometries. Furthermore, comparison with the experiment becomes progressively more complex. These challenges are obvious even in theory, where the analysis of interaction and structure evolution in systems of non-spherical particles evolved much slower than for spherical ones. These difficulties are progressively magnified for the experimental systems, characterized by the presence of the particle size distributions, noise, *etc*.

Here we introduce the approach for the analysis of the structure evolution in the system of interacting anisotropic particles based on a combination of deep fully convolutional neural networks and variational autoencoders (VAEs) with rotational invariance. This methodology particularly relies on the recognized capability of the VAEs to reduce high-dimensional data sets to the low-dimensional continuous latent variables, and disentangle the representations, *i.e.* discover the significant trends in data.[38-42] The examples of such trends are writing styles in hand-written digit data bases or emotional states in human face databases. Introduction of the rotation angle in the image plane as one of the latent variables allows identification of the variants of the same shape at multiple orientations. In this manner, we introduce the parsimonious particle level descriptors for the system. The time dynamics of the global averages provide the insight into the system evolution, emergence of order, and phase transitions, whereas trajectory-level analysis yields insight into particle dynamics. Note that while this approach is illustrated here for a specific case of particle self-assembly visualized *via* liquid atomic force microscopy, it is in fact universal and can be applied for the analysis of structure evolutions across broad experimental and theoretical domains.

**Results and Discussion**

As a model system, we explore the self-assembly of a *de novo* designed helical repeat protein DHR10-mica18 (where 18 indicates the number of repeats of the protein),[43] on muscovite mica (*m*-mica) which was recorded by high-speed atomic force microscopy (HS-AFM). Inspired by how antifreeze proteins bind ice crystal surfaces, DHR10-mica18 geometrically matches the potassium ($K^+$) sublattice on the (001) plane of m-mica with a designed interface containing 54 carboxylate residues. Its large size and aspect ratio allows individual proteins to be resolved as they adsorb to the mica surface and assemble into ordered structures that are modulated by the concentration of salt in solution.[43] With 100 mM KCl, DHR10-mica18 self-assembles into numerous discrete domains aligned along one of the three closed-packed $K^+$ lattice-directions. However, the close-packed matrix does not have a steady state, instead the domains fluctuate in both direction and size and hence the relative positions of individual proteins within the domains remain dynamic as well, Figure 1. Tracking DHR-10-mica18 on m-mica with *in-situ* HS-AFM generates a large amount of data (hundreds of frames with tens of protein molecules in each frame), that is beyond the capability of manual approaches of statistical analysis to define descriptors of the self-assembly process. Here, we develop a machine learning approach to explore this process.

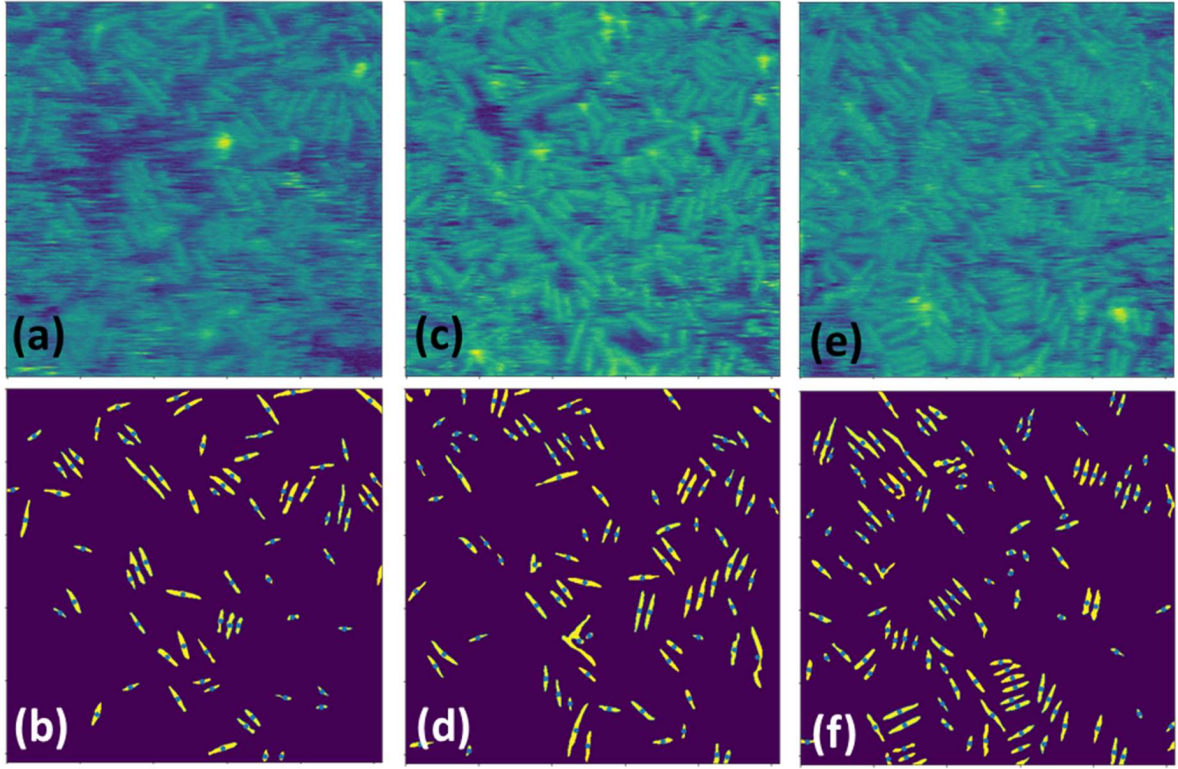

**Figure 1.** The AFM studies of the dynamics of protein organization. Shown are raw AFM images (top row) and neural network reconstructions (bottom row) for frames (a,b) +131.6 s, (c,d) +263.2 s and (e,f) +394.7 s. The dots in (b,d,f) mark the positions of the centers of gravity of the particles. Image size is 200 nm and frame rate is 0.38 Hz.

To analyze particle positions and orientations visualized in AFM data, we first use the deep convolutional neural network (DCNN) approach to remove noise and identify particles centers of the mass, size, and orientations *via* semantic segmentation (*i.e.,* identifying each pixel in the image as belonging to the particle or substarte). Previously, this approach has been used for analysis of atomically resolved imaging in Scanning Transmission Electron Microscopy[44, 45] and Scanning Tunneling Microscopy,[46] Scanning Electron Microscopy,[47-52] Transmission Electron Microscopy,[50, 53-56] Scanning Probe Microscopy,[57-59] Optical microscopy,[60-63] and Transmission X-ray microscopy.[64, 65] Here, a training set for DCNN is made from the sub-set of original images with clearly discernible particles as features and manually labeled categorical images (particle/not particle) as a target. The training set is augmented *via* rotations, Gaussian noise, and horizontal or vertical flips.

Generally, the U-Net[66] and its various extensions (Dense-U-Net,[67] U-Net++,[68] Res-U-Net,[69] *etc.*) are considered as the go-to DCNN models for semantic segmentation problems. However, we found that these models may not always properly segment particles that form domains (*i.e.*, are located close to each other) in noisy AFM data, which is likely caused by the information loss during the multiple down-sampling/pooling (and corresponding up-sampling/un-pooling) operations. One solution is to reduce the number of blocks in U-Net (and hence, the number of pooling/un-pooling operations), while increasing the depth of the bottleneck layer and replacing the regular convolutions in the bottleneck layer with cascades of dilated convolutions with different dilation rates. The dilated convolutions help maintaining the ability to recognize features at different scales and the computational feasibility. Here, we use the in-house developed *dilnet* neural network architecture with only one max-pooling (and the corresponding un-pooling operation) to preserve the maximum amount of information and two cascades of dilated convolutions with dilation rates of 2, 4, and 6 in the bottleneck layer. To further improve the network's predictions, we utilize the deep ensemble approach, where multiple networks with different "training trajectories" are trained in parallel and used for prediction. The ensemble mean prediction and the associated variance provide improved recognition/generalization and uncertainty estimates.[70] Note that thus trained network ensemble yields the semantically segmented images, *i.e.* for each pixel in the raw image Fig. 1 (a,c,e) the decoded image Fig. 1 (b,d,f) yields the "probability" that it belongs to the particle.

With this information in hand, the positions of the center of mass of individual particles can be readily mapped. Previously, we used the ellipsoid fit to find the characteristic particle sizes and rotation angles. Based on these top-down particle level descriptors, the system evolution in terms of relevant distribution functions and their spatial correlations can be analyzed. In certain cases when the same particle can be traced across multiple frames, the evolution of these parameters along the trajectory can be traced. Note that while these characteristics are readily available in the simulation studies, discovery of these in the experimental data represents a considerable challenge and analysis is naturally limited to the objects that can be discerned by the deep learning network and, for trajectories, the displacements of which is sufficiently small to allow for reconstruction. That said, the subsequent discussion and analysis can be equally applied for both decoded experimental and simulation data.

To get insight into particle dynamic and structure evolution during the self-organization process, here we use a rotationally invariant extension[71] of the variational autoencoder (rVAE). Generally, autoencoders (AE) refer to the class of the neural networks that compress the data set to a small number of latent features, and then expand back to original data set. The training aims to minimize information loss between the initial and reconstructed images *via* usual backpropagation. This process tends to select the relevant features in the data set and reject the noise, giving rise to applications for denoising, *etc*. At the same time, the latent features allow for efficient encoding of the original data set.

Variational autoencoders (VAEs) expand on this concept by substituting the bottleneck latent layer by the latent space, from which the variables for decoding are drawn from a prescribed prior distribution. In this manner, VAEs represent a hybrid of the AE approach for creating generalized encoding and decoding functions, and Bayesian priors for feature selection. Since the Bayesian layer is non-differentiable, the training of the VAEs is based on sampling latent vector using the reparameterization trick as shown by Kingma and Welling.[38, 39] The unique aspect of VAEs is that they tend to structure the latent space in such a way that the decoded data will have clear variations along the latent directions. This behavior, referred to as disentangled representations,[72, 73] allows for determination of styles of handwriting of fashion, style transfer, *etc*. More generally, VAEs allow projection of high-dimensional and potentially discrete spaces onto low-dimensional differentiable manifold, potentially allowing for mapping equations of motions, enabling Bayesian optimization, *etc*.

Here, we adapt the variational autoencoder to include the rotation and offsets in *x*- and *y*-directions as three of the latent variables, in addition to classical latent variables. In this manner, the rotations of the particles in the image plane are separated as one latent variable, and non-idealities in determination of particle center of mass are captured *via* offsets. The remaining latent variables provide the information on particle shape, structure of the nearest neighborhood, *etc*. depending on the size of the sampling window (size of sub-image cropped around each detected particle). The encoder and decoder of rVAE are chosen to be simple fully connected ("dense") neural networks.

The semantically segmented DCNN output is used to create an input into the rVAE. The use of the raw data led to relatively smooth decoded features that allow for partial orientation mapping only and was not actively pursued. The latent angles (in this specific case) give rise to clear multimodal distributions corresponding to 6 possible orientations of the particles on the surface, whereas offset distributions are reasonably narrow and sharp. These criteria were used to identify optimal training and sampling window parameters. Here, we have chosen the windows slightly above the particle length (see Fig. 2d-f), to fully capture the particle shape and the lateral interactions but at the same time avoid excessive details that necessitate high dimensionalities of latent space to describe.

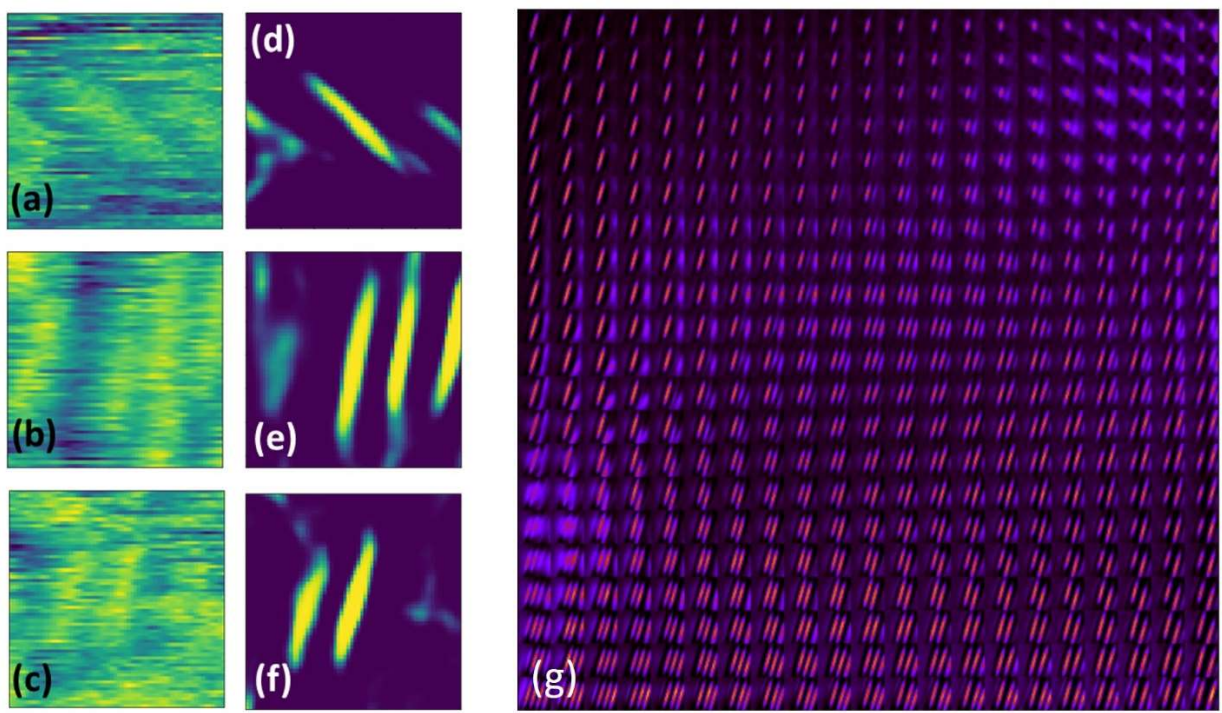

**Figure 2**. (a,b,c) Several sub-images of the raw data and (d,e,f) corresponding DCNN output. The latter is used for rVAE training. (g) The learned latent space of the rVAE projected onto the image space showing the evolution of the decoded images as a function of latent variables, $d$ ($L_1$, $L_2$). The size of sub-images is 25 nm.

The rVAE's training was performed using Adam optimizer[74] for 3000 epochs with a mini-batch size of 200 for the data set of the ~13,200 sub-images, as limited by the number of the decoded particles in the data set. The typical sub-images with raw experimental data and corresponding decoded images are shown in Figure 2. rVAE analysis converts each decoded sub-image into the latent angle ($L_\theta$), offsets ($L_{\Delta x}$ and $L_{\Delta y}$), and latent variables associated with particles structure and spatial arrangement ($L_1$ and $L_2$) that now describe the state of each particle. The system behavior can then be analyzed *via* statistical analysis of the time dependence of relevant distributions within each frame as a function of frame number, correlation function analysis, or trajectory analysis.

A convenient way to represent the rVAE operation is through the analysis of latent space representations. The encoding of the sub-images transforms each of the sub-images into 3+2 latent variables. The distribution of the latent variables determines the size of the latent variable space bound by minimal and maximal values of $L_1$ and $L_2$. We can further introduce the uniform rectangular grid of points in the latent space and decode these values to yield the particles geometry. This latent space representation is shown in Fig. 2 (g). Note that the decoded features have a clear physical meaning, representing the single and multiple particles assemblies.

The characteristic aspect of VAEs is their potential for disentangling the data representations, where each latent variable describes the certain trait in data set. This examination of this behavior is visible in Fig. 2 (g), where on transition from left to right (*i.e.*, for constant $L_2$) the particles become smaller, whereas on vertical the number of particles in decoded sub-image increases. The projected latent space for the vanilla VAE can be found in the Supplemental Material.

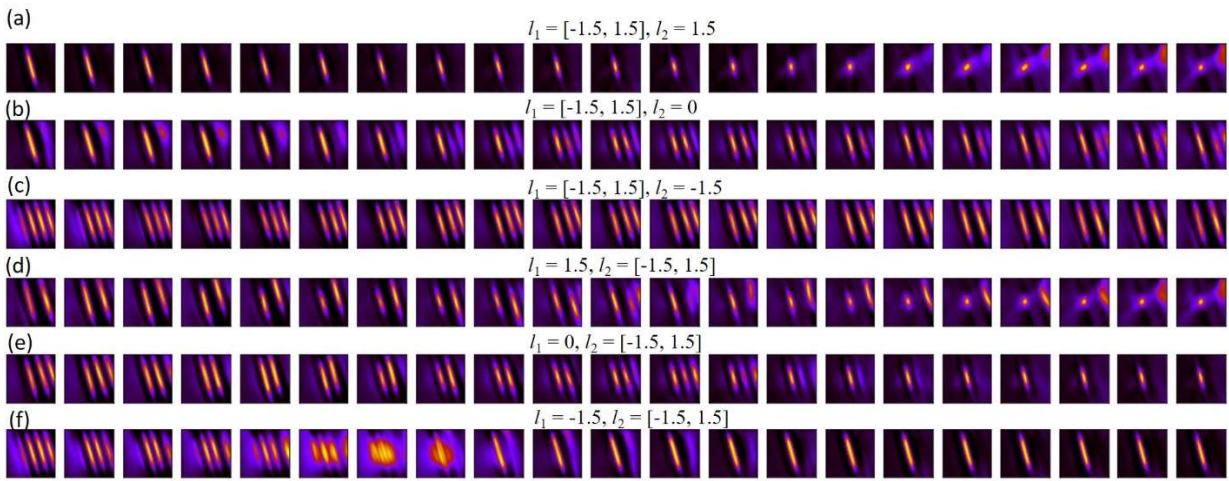

**Figure 3.** (a,b,c) Magnified cross-sections along the $L_1$ axis at fixed values of $L_2$ and (d,e,f) along the $L_2$ axis along the fixed values of $L_1$.

To get further insight into these behaviors, shown in Fig. 3 are the expanded cross-sections from Figure 2 (g) obtained for higher sampling density in latent space. Here, the Fig. 3 (a) shows the gradual evolution of a particle shape, eventually converting to non-physical mixed contrast. Similar variation between physical and unphysical shapes is observed in (d). At the same time, Fig. 3 (b,c,e,f) illustrate the evolution of density of particles across the row.

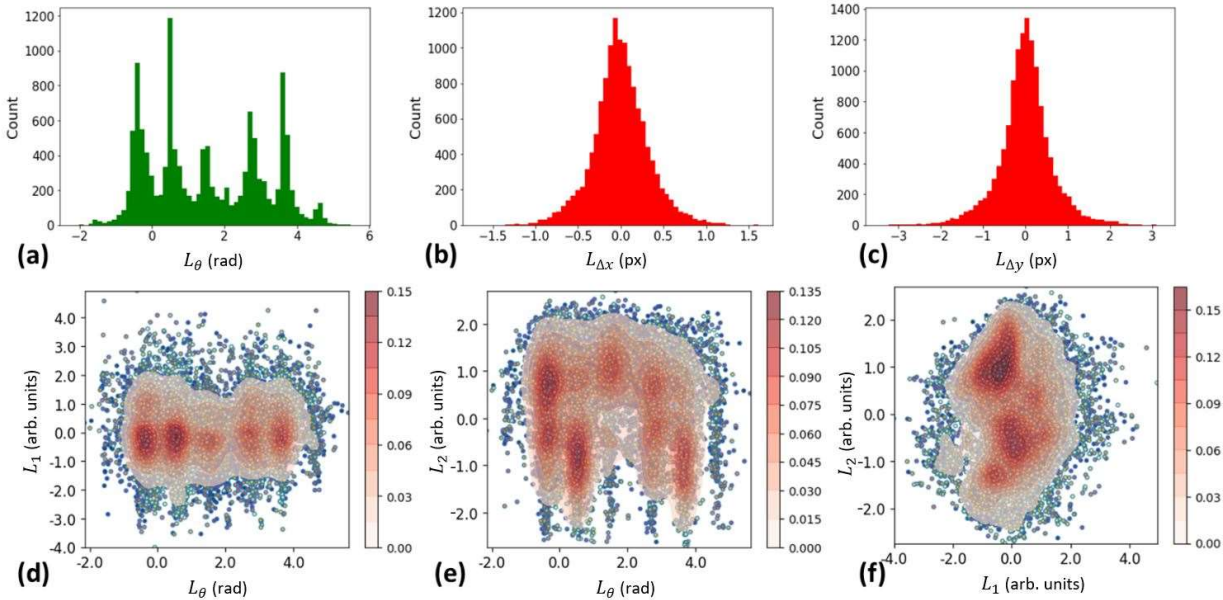

**Figure 4.** Distributions of latent variables across the full experiment. (a) Distribution of the encoded angle ($L_\theta$) clearly illustrates the presence of six dominant orientations due to the interactions between the nanoparticles and substrate. Note that here particles are not assumed to be symmetric perpendicular to the long axis. (b, c) Distributions of the *x*- and *y*-offsets ($L_{\Delta x}$ and $L_{\Delta y}$). These distributions are relatively featureless but indicate the convergence of the rVAE. (d,e,f) Joint distributions between the encoded angle $L_\theta$ and latent variables $L_1$, $L_2$ associated with particles structure/arrangement. Shown are the points corresponding to each particle and superimposed is kernel density estimate.

The latent representation analysis allows us to get further insight into the global system dynamics *via* the analysis of the latent parameter distributions and their time dynamics. Shown in Figure 4 is the global (*i.e.* averaged over the full data set) distributions for the latent variables. Here, the angle distributions clearly show 6 peaks, corresponding to preferential orientation of the protein particles due to anisotropic interactions with the substrate. The offset distributions are relatively featureless and generally confined within one pixel, indicative of successful particle finding (in cases when rVAE fails to converge, much broader distributions are observed). Finally, the joint distributions in latent variable space are shown in Figs. 4 (d-f). Here, we visualize both the individual data points and the superimposed kernel density estimates. Note that the $L_1$-$L_\theta$ distribution is almost marginalizable, with clear 6-fold maxima associated with angle distribution. At the same time, the $L_2$-$L_\theta$ distribution is nontrivial, illustrating the presence of the multimodal $L_2$ distributions for each angle. Comparison with Fig. 2 (g) illustrates that these distributions differ by the lateral spacing between the particles. This behavior is further illustrated in Fig 4 (f), where two primary maxima corresponding to ~1 and -1 for $L_2$ are clearly seen.

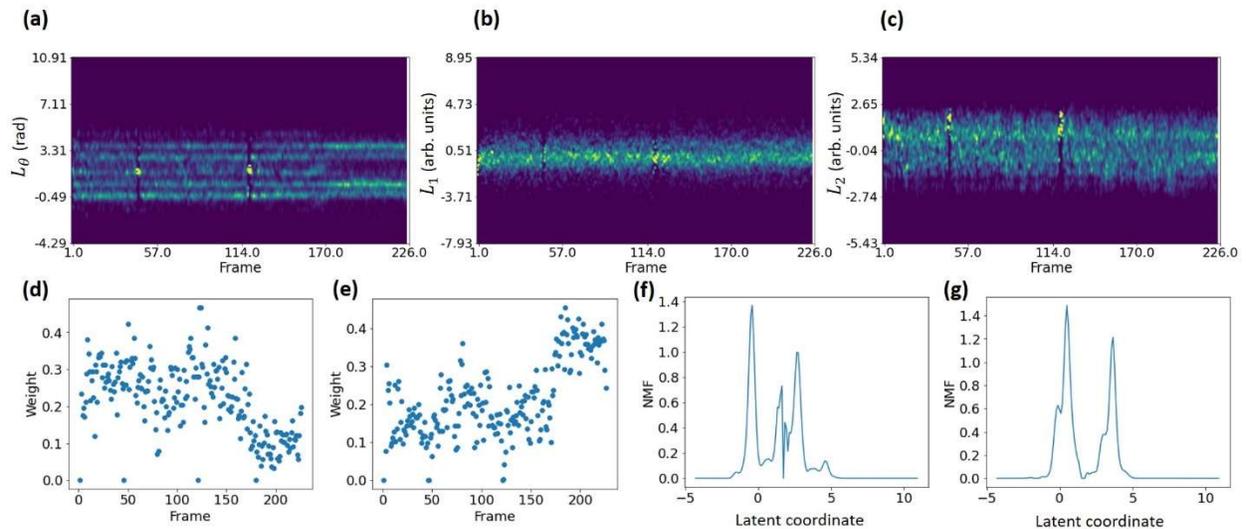

**Figure 5.** Time dynamics of (a) encoded angle $L_\theta$, (b) first ($L_1$) and (c) second ($L_2$) latent variables. (d,e) time dependence of the weights of the first and second NMF component and (f,g) corresponding endmembers. Note that features with lost contrast were excluded from NMF analysis. Clear transitions at frame ~170 is visible (a,d).

To get insight into the dynamics of the system, we analyze the time dynamics of the latent variable distribution. To accomplish it, we calculate the 1D kernel density estimates (KDE) for the distribution of the corresponding latent variables, as shown in Fig. 5. The time dynamics of the angle clearly indicates the presence of six rotational variants in the early times, with the transition to only 4 dominant variants on later stages.

The relevant aspects of this behavior can be further analyzed using the suitable dimensionality reduction method. Given that KDE are positively defined, we use the non-negative matrix factorization (NMF) such that a separation of the 2D data set KDE($L_c$,t), where $L_c$ is a chosen latent variable and $t$ is time, is represented as

$$KDE(c,t) = \sum_{i=1}^{N} w_i(L_c) e_i(t) \qquad (1)$$

where $w_i(L_c)$ are the NMF weights and $e_i(t)$ are the endmembers that determine characteristic time behaviors. The number of components $N$ is set at the beginning of the analysis and can be chosen based on the quality of decomposition, anticipated physics of the systems, *etc*. Here, after experimentation, the $N = 2$ was found to be sufficient to represent the observed dynamics.

The time dependence of the 1st and 2nd NMF component are shown in Figure 5 (d,e) respectively. Note the sharp change at frame ~140 associated with the disappearance of 2 out of 6 orientational variants, *i.e.* spontaneous symmetry lowering in the system.

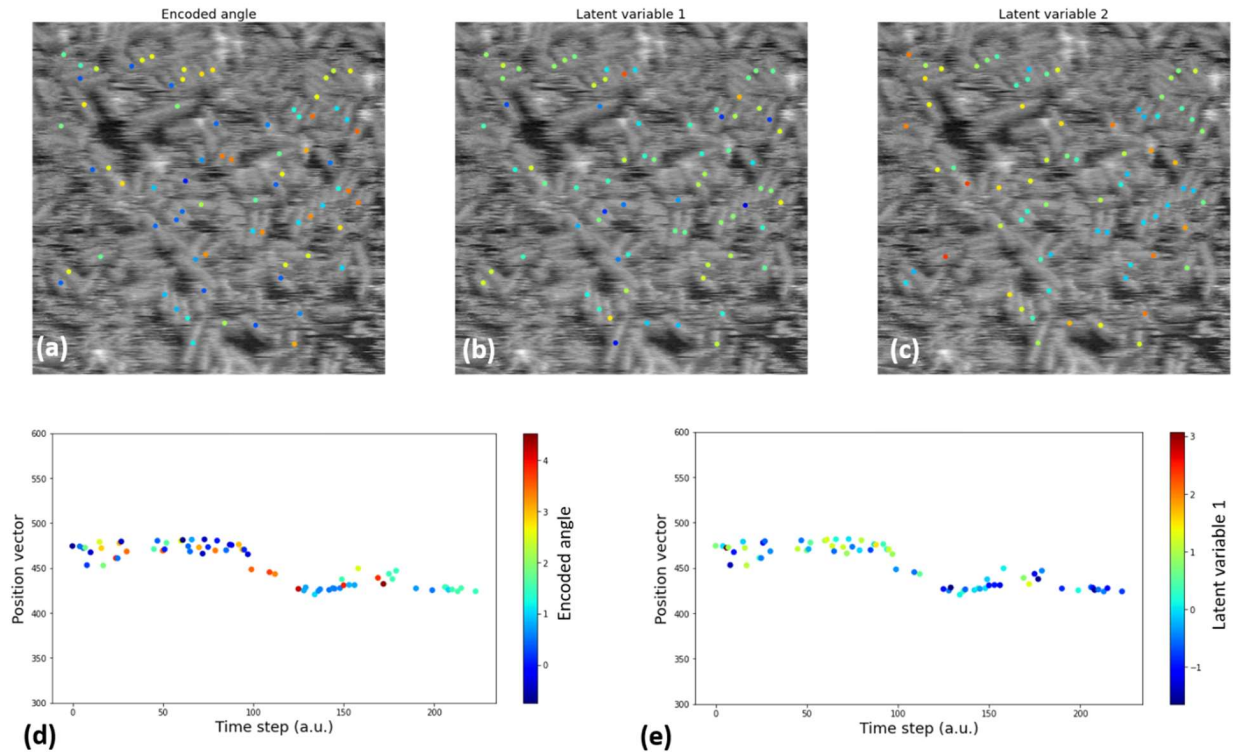

**Figure 6.** Latent encoding in particle assemblies. (a) Latent angle and (b) first and (c) second latent variables. Note that for ordered phase both parallel (ferroelectric-like) and antiparallel

(antiferroelectric-like) arrangements of the particles can be observed. (d, e) Example of a trajectory of a single particle encoded by angle (d) and the first latent variable (e). Image size is 200 nm.

Finally, the latent representations allow the exploration of the dynamics on a single particle level. Shown in Figure 6 are the raw images at several time steps with the color markers indicative of the corresponding latent variables. In some cases, the particle trajectories can be reconstructed by tracking individual particles from one frame to another *via* the nearest neighbor search. Here the search radius was set to 21 px, which corresponds to 8.2 nm. In this case, the evolution of the latent variable along the trajectory can be explored, as shown in Fig. 6 (d,e). Similarly, the particle dynamics can be explored in the latent space, representing the changes in the particle geometry and nearest neighborhood during the evolution.

**Conclusions**

To summarize, the particle dynamics during the protein self-assembly was explored using machine learning workflow that combined DCNN based semantic segmentation and rotationally invariant VAE analysis of orientation shad shape dynamics. The chosen VAE architecture allows disentangling the particle orientation from other degrees of freedom and compensates for shifts. The disentangled representations in the latent space encode the rich spectrum of local transitions that can now be visualized and explore *via* continuous variables. The time dependence of ensemble averages allows insight into the time dynamics of the system, and in particular illustrates the presence of the potential ordering transition. Finally, analysis of the latent variables along the single particle trajectory allows mapping evolution of particle shape and nearest environment. This approach allowed us to detect the symmetry lowering transitions associated with emergence of long-range order within the structures. More generally, this approach now allows for straightforward analysis of distributions and fluctuations of ordered domains, tied to relative free energies of the various states.

The proposed approach is expected to be universally applicable for the description of the imaging data in optical, scanning probe, and electron microscopy seeking to understand dynamics of complex systems where rotations is a significant part of the process. We note that while here both the DCNN and rVAE were applied to a single class, it can be expanded to multiclass features in a straightforward way. Furthermore, this approach can be used for exploration of computationally generated datasets, including the evolution of the electronic density and lattice displacements desiring diffusion and reactions in atomistic modelling, dynamics of macromolecules, *etc*. Finally, it directly allows to enumerate and explore transformation mechanisms in system with complex interacting objects *via* analysis of corresponding latent representations.

**Methods:**

**Particle synthesis**
DHR10-mica18 proteins were expressed in *E. coli*, purified with nickel NTA affinity and size exclusion chromatography, and dialyzed into 20mM Tris buffer (pH=8). For more details see reference [43].

**High-speed atomic force microscopy**
DHR10-mica18 protein stock solution was diluted to 0.025 μM with 20 mM Tris buffer (pH=7) having 100 mM KCl. 20 μl diluted protein solution was dropped onto freshly cleaved muscovite-mica (001) (SPI Supplies) and characterized by Cypher Video-Rate AFM (Asylum Research) in liquid amplitude-modulation mode. The probe USC-F1.2-K0.15 (NanoWorld) was used. The imaging force was adjusted to minimize any interruption. Tris-HCl buffer (pH 7, 1 M), and KCl were bought from Sigma-Aldrich. Nuclease-free water was bought from Ambion.

**Data analysis**
The DCNN and rVAE were implemented *via* AtomAI package.[75] To train a deep DCNN ensemble we first trained a baseline model for *N* epochs until test loss reached a plateau. We then trained 12 individual ensemble models for *n* epoch ($n \ll N$) starting each time with the weights of the baseline model and performing training data shuffling with different random seed. The Adam optimizer[74] with a learning rate of 0.001 was used for optimizing weights of all the ensemble models. The encoder and decoder of rVAE were 2-layer perceptrons with 128 neurons in each layer activated by *tanh()* function whose weights were optimized using Adam optimizer with a learning rate of 0.0001. We have experimented both with the convolutional and fully-connected in the rVAE's encoder and decoder, and for this data did not see significant difference in performance. The latent space consisted of 5 neurons: one neuron designated to absorb particles rotation, two other neurons designated to absorb particles translation, and the remaining two neurons for capturing the remaining factors of variations (*i.e.*, particles structure). We note that while more neurons can be in principle added to analyze variations in particles structure, it complicates the visualization of the learned latent manifold and has not shown to lead to any additional new insights on this data set.

**Supporting Information**

The supporting information is available free of charge on the ACS Publication website at DOI: xxxx.

Supplementary figure showing the learned latent manifold of a regular VAE model

**Acknowledgment**

This research including AFM (S.Z., J.J.D.,), data analytics (S.V.K.) and protein design and synthesis (H.P., D.B.) are supported by the US Department of Energy, Office of Science, Office of Basic Energy Sciences, as part of the Energy Frontier Research Centers program: CSSAS–The Center for the Science of Synthesis Across Scales–under Award Number DE-SC0019288, located at University of Washington. The machine learning was performed and partially supported (M.Z.) at the Oak Ridge National Laboratory's Center for Nanophase Materials Sciences (CNMS), a U.S. Department of Energy, Office of Science User Facility. High-speed AFM experiments were performed at the Department of Energy's Pacific Northwest National Laboratory (PNNL). PNNL is a multi-program national laboratory operated for Department of Energy by Battelle under Contract No. DE-AC05-76RL01830.

**TOC**

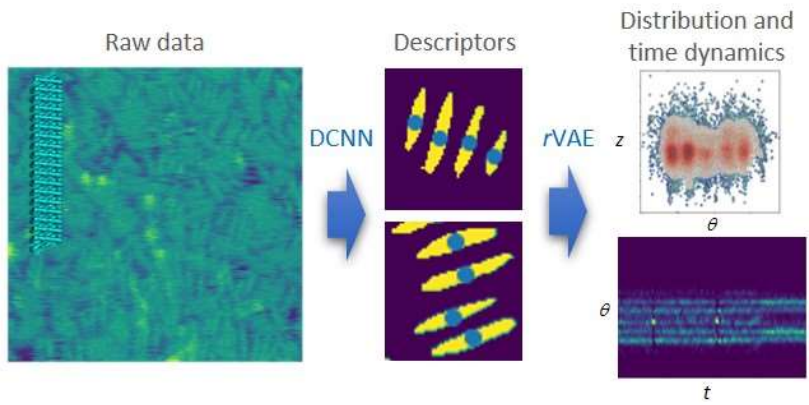